  \documentclass[prl,aps,twocolumn,showpacs]{revtex4}
\usepackage[dvips]{graphicx}
\usepackage{dcolumn}%
\usepackage{amsmath}%
\usepackage{amsfonts}%
\usepackage{amssymb}
 \usepackage[usenames]{color}
\providecommand{\U}[1]{\protect\rule{.1in}{.1in}}
\newcommand{\be}{\begin{equation}}
\newcommand{\ee}{\end{equation}}
\newcommand{\bea}{\begin{eqnarray}}
\newcommand{\eea}{\end{eqnarray}}
\newcommand{\bt} {\begin{tabular}}
\newcommand{\et} {\end{tabular}}
\newcommand{\nn}{ \nonumber}
\newcommand{\ds}{\displaystyle}
\newcommand{\ba} {\begin{array}}
\newcommand{\ea} {\end{array}}
\begin{document}

\title{Thermally induced charge current through long molecules}

\author{  Natalya A. Zimbovskaya$^1$ and Abraham Nitzan$^{2,3}$}

\affiliation
{$^1$Department of Physics and Electronics, University of Puerto 
Rico,  Humacao, PR 00791, USA}  

\affiliation{$^2$Department of Chemistry, University of Pennsylvania, Philadelphia, PA19104, USA}

\affiliation{$^3$School of Chemistry, Tel Aviv University, Tel Aviv, Israel}

\begin{abstract}
In this  work we theoretically study steady state thermoelectric transport through a single-molecule junction with a long chain-like bridge. Electron transmission through the system is computed using a tight-binding model for the bridge. We analyze dependences of thermocurrent on the bridge length in unbiased and biased systems operating within and beyond linear response regime. It is shown that length-dependent thermocurrent is controlled by the lineshape of electron transmission in the interval corresponding to HOMO/LUMO transport channel. Also, it is demonstrated that electron interactions with molecular vibrations may significantly affect length-dependent thermocurrent.
		\end{abstract}


\date{\today}
\maketitle

\section{I. Introduction}

Presently, charge transport through molecular systems is an important research field because of possible applications of these systems in molecular electronics \cite{1,2,3,4}. The key element and  basic building block of molecular electronic devices is a single-molecule junction including a couple of metallic/semiconducting electrodes linked by a molecular bridge.  Alongside  other tailored nanoscale systems (such as carbon-based nanostructures and quantum dots),  single-molecule junctions hold promise for enhanced efficiency of heat-to-electric energy conversion \cite{5,6,7,8,9,10,11,12}. Therefore, thermoelectric properties of single-molecule junctions are being explored both theoretically and experimentally. 

In general, thermoelectric charge transport through molecular junctions is controlled by  simultaneous driving by electric and thermal driving forces. The combined effect of these forces depends on several factors including the bridge geometry and  the characteristics of its coupling to the leads \cite{13,14,15,16,17,18,19,20,21,22,23,24} and electron-electron interactions \cite{25,26,27,28,29,30,31,32,33,34}.  Thermoelectric transport characteristics may be affected due to 
 interaction between transmitting electrons and environmental nuclear motions \cite{35,36,37,38,39,40,41,42,43,44,45,46,47,48,49,50,51,52,53,54} and  the effects of quantum interference  \cite{55,56}.  Electron-photon interactions may also bring changes into thermoelectric properties of nanoscale systems \cite{57}.  Under certain conditions (e.g. in single-molecule junctions with ferromagnetic electrodes and/or with a magnetic molecule used as a linker) spin polarization of electrons may significantly influence thermoelectric transport resulting in several new phenomena such as spin Seebeck effect  \cite{58,59,60,61,62,63,64,65,66,67,68,69}.

It was repeatedly demonstrated that electron transport through molecules strongly depends on the molecular length. 
  Length-dependent electronic conductance as well as the electronic contribution to heat conductance and thermoelectric response were observed and discussed in molecular junctions with repeating molecular  units such as benzene or phenyl rings 
\cite{13,14,15,16,17,18,19,20,21,22,23,70,71}. These linkers provide a better opportunity to observe  relationships between transport coefficients and the length of the linker. For other kinds of molecular bridges these relationships are less distinct due to the diversity of specific properties of different parts of the bridge. 
Usually, these transport characteristics are measured assuming that thermal gradient $ \Delta \theta $ applied across the system is much smaller than the average temperature $ \theta $ characterizing the latter  \cite{72}.  In this case, thermal driving forces remain relatively weak and the response of the system is linear in $ \Delta \theta. $ Correspondingly, transport characteristics such as electron conductance and thermopower
appear to be  independent on $ \Delta \theta. $ However, as the temperature gradient $ \Delta \theta $ increases, the system may switch to a regime of operation nonlinear in $ \Delta \theta. $ In several recent works nonlinear thermoelectric transport through molecular junctions and other tailored nanoscale systems has been discussed \cite{35,37,43,49,58,73,74,75,76}. Nonlinear Seebeck effect was already observed in semiconducting quantum dots and single molecule junctions  \cite{19,77,78}. 

An important characteristic of thermoelectric transport through molecular junctions and similar nanoscale systems is  the thermocurrent, defined as a difference between the charge current flowing through a biased system in the presence of a temperature gradient $ \Delta \theta $ and the current flowing through the same system when the temperature gradient is removed $(\Delta \theta = 0) $ \cite{77}. According to this definition, $ I_{th} $ represents the contribution to the charge current which appears due to thermally excited transport of charge carriers through a molecular junction. The thermocurrent is more convenient for measuring and modeling than some other characteristics of thermoelectric transport such as the Seebeck coefficient. Also, it may be studied within the same computational approach over a wide range of  $ \Delta \theta $ values both within and beyond the linear response regime. Despite these  advantages, the thermocurrent properties in nanoscale systems were not so far  thoroughly studied. 
 In accord with the well-pronounced dependences of both electron tunnel conductance and thermopower on the length of a molecular bridge  we  expect $ I_{th} $  to be  length-dependent as well. Also, we may expect that molecular vibrations can significantly affect thermocurrent.  In the present work we focus on  this observable and its dependences on the molecular linker length and on the effect of vibrational phonons.

\section{ii. model and main equations}

In the following analysis we assume coherent electron  transmission
 to be  the predominant transport mechanism. We simulate a linker in a single-molecule junction by a periodical chain of $ N $ sites. Each single site is assigned   an  on-site energy $ E_i $ and coupled to its nearest neighbors with the coupling strengths $ \beta_{i-1,i} $ and $ \beta_{i,i+1} ,$ respectively $( 2 \leq i \leq N-1 ). $ Within the simplest version of this model, all on-site energies are assumed to be equal $(E_i = E_0) $ as well as all coupling parameters $(\beta_{i-1,i} = \beta_{i,i+1} = \beta). $  Such simple chain models  are often used to represent molecular bridges  comprising repeating units  where $ \pi - \pi $  coupling dominates electron transport  and  the parameter $ \beta $  characterizes the coupling between adjacent $ \pi $ orbitals  \cite{79}. The schematics of this model showing relevant parameters employed  in further computations is presented in Fig. 1. 
 This simple chain model may be modified by introducing gateway states different from the rest. This could be achieved by separating out two sites at the ends of the chain, setting on these terminal sites on-site energies $ E_i = E_N = \epsilon $ which differ from $ E_0 $ and suggesting that these sites are coupled  to their neighbors with different strength$(\beta_{1,2} = \beta_{N-1,N} = \delta). $ It was shown that gateway states may significantly affect  the length dependence of the Seebeck coefficient \cite{19,80} and  similar effects  may appear in the thermocurrent.   Such effects can be investigated with the simple model advanced here.

\begin{figure}[t] 
\begin{center}
\includegraphics[width=7cm,height=6cm]{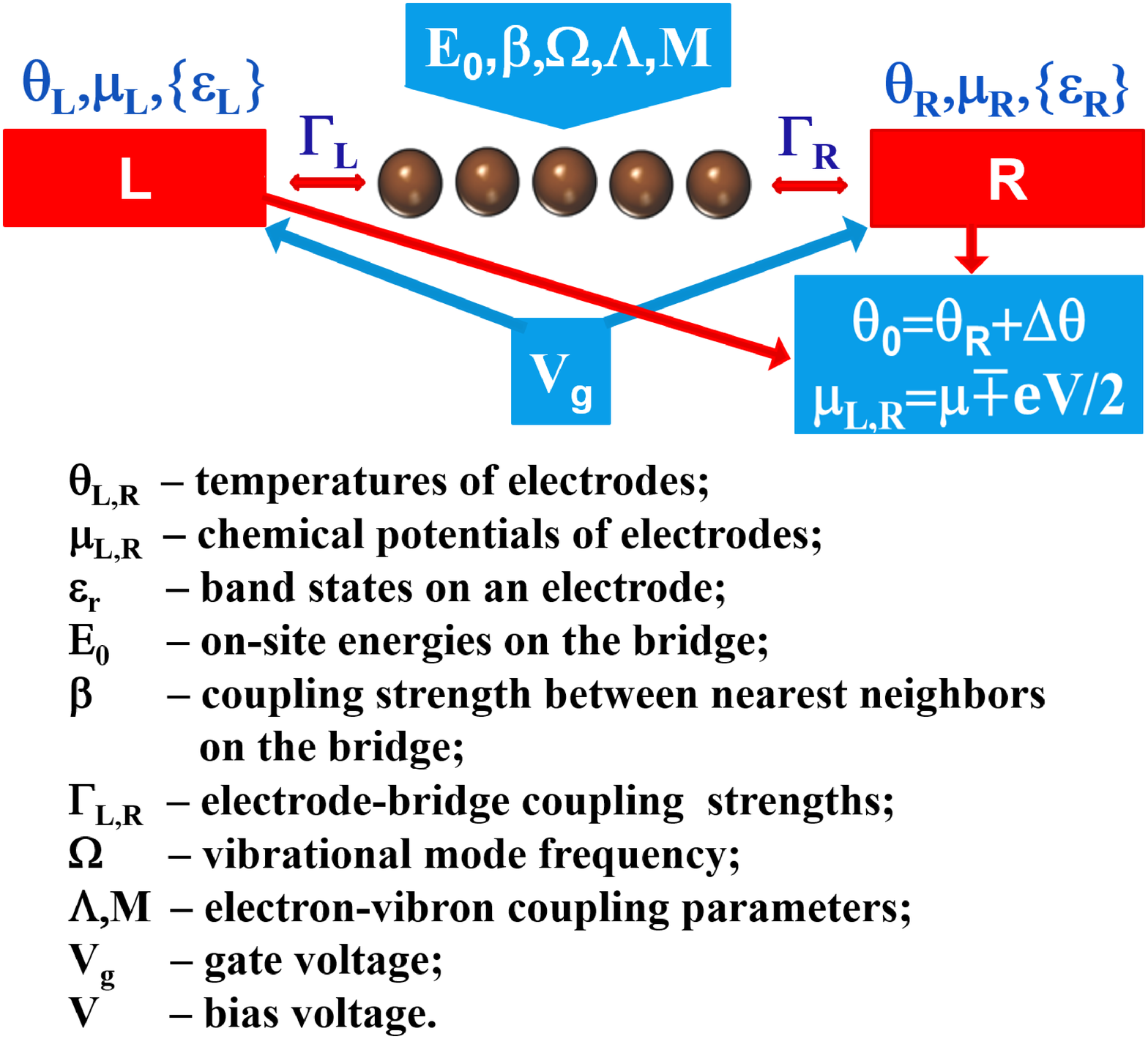}
\caption{(Color online) Schematics of the metal-molecule-metal junction used to analyze thermally induced transport. Indicated parameters are relevant energies determining the behavior of the themocurrent.
}
 \label{rateI}
\end{center}\end{figure}

The Hamiltonian of a single-molecule junction where the chain-like bridge interacts with a single vibrational mode may be written as 
\be
 H = H_M + H_L + H_R + H_T + H_{ph}.  \label{1}
\ee    
For a simple chain, the term$ H_M $  that represents the molecular bridge coupled to the phonon mode has the form:
\begin{align}  
H_M = &  E_0\sum_{i,\sigma} d_{i\sigma}^\dag d_{i \sigma} + \beta \sum_{i, \sigma}   d_{i \sigma}^\dag \big[d_{i+1, \sigma} (1 - \delta_{iN})  
\nn\\  & +   
d_{i-1, \sigma} (1 - \delta_{i1}) \big] + 
\sum_{i j \sigma} \Lambda_{ij} d_{i \sigma}^\dag d_{j \sigma} (a^\dag + a).          \label{2}
\end{align}
Here, $    1 \leq i,j \leq N,\   d_{i\sigma}^\dag,\ d_{i\sigma} $ are creation and annihilation operators for electrons with the spin $ \sigma $ on the bridge site $"i",\ a^\dag, \ a $ are creation and annihilation operators for the phonon mode and $ \delta_{ik} $ is the Kronecker symbol. The last term in expression (\ref{2}) describes electron-vibron interaction.  In the following analysis we assume that the coupling parameters accept nonzero values  only when $ i = j\ (\Lambda_{ii} \equiv \Lambda) $ and when $ i = j  \pm 1 \ (\Lambda_{i, i+1} = \Lambda_{i, i-1} \equiv M) $. It is reasonable to postulate that $ M < \Lambda.$  

The terms $ H_\gamma \ (\gamma = L,R) $  correspond to noninteracting electrons on the leads with energies $ \epsilon_{r\gamma\sigma}: $
\be
H_\gamma = \sum_{r,\sigma} \epsilon_{r\gamma\sigma} c_{r\gamma\sigma}^\dag 
c_{r\gamma\sigma}    \label{3}
\ee
where $ c_{r\gamma\sigma}^\dag  $ and $ c_{r\gamma\sigma} $ create and annihilate electrons on the leads. Within the employed model, only terminal sites of the chain are coupled to electrodes, so the transfer Hamiltonian $ H_T $ can be written as:
\be
H_T = \sum_{r\sigma} \big(\tau_{rL\sigma} c_{rL\sigma}^\dag d_{1\sigma} + 
\tau_{rR\sigma} c_{rR\sigma}^\dag d_{N\sigma} \big) + H.C. \label{4} 
\ee
This term describes electron tunneling between the bridge and the electrodes,  where factors $ \tau_{r\gamma\sigma} $ characterize the coupling of relevant electron states on the bridge to those on the leads. Finally, the term $ H_{ph} $ represents  the vibrational mode with the frequency $ \Omega: $ 
\be
H_{ph} = \hbar\Omega a^\dag a.  \label {5} 
\ee  
We note that the last term in the expression (\ref{2}) may be simplified by assuming that electron-phonon coupling parameters $ \Lambda_{ii} $  take on the same value for all electron states on the bridge. However, this simplification is fitting well with the assumption concerning on-site energies, so we use it in following calculations. As follows from Eq. (\ref{2}) interactions between electrons on the bridge are omitted from consideration. More advanced models can be used but the simple model (\ref{2}) already shows the essential physics that needs to be addressed.

To  eliminate the electron-phonon coupling term from the Hamiltonian (\ref{1}) we employ the commonly used small polaron (Lang and Firsov) transformation which converts the Hamiltonian $ H_M $ into $ \tilde H_M = \exp[s] H_M \exp[-s] $ where $  \ds s = \frac{\Lambda}{\hbar \Omega}\sum_{i,\sigma}d_{i \sigma}^\dag d_{i \sigma} (a^\dag - a)  +  \frac{M}{\hbar \Omega} 
 \sum_{i,\sigma} d_{i \sigma}^\dag \big[d_{i+1, \sigma} (1 - \delta_{iN})  + d_{i-1, \sigma} (1 - \delta_{i1}) \big] (a^\dag - a)  $   \cite{81,82}.  As a result, we obtain:
\begin{align}
\tilde H_M = & \tilde E_0 \sum_{i,\sigma} d_{i \sigma}^\dag d_{i \sigma}   \label{6}
\\ & + 
\tilde \beta \sum_{i,\sigma} d_{i \sigma}^\dag \big[d_{i+1,\sigma} (1 - \delta_{iN}) + d_{i-1,\sigma}  (1 - \delta_{i1}) \big] . \nn
\end{align}
Here, the on-site energy acquires a polaronic shift  originating from electron-vibron interaction: $ \tilde E_0 = E_0 - \Lambda^2/\hbar\Omega. $  Assuming a sufficiently weak electron-phonon coupling $(\Lambda \ll \beta) $ the coupling parameter $ \beta $ is renormalized in a similar way: $ \tilde \beta = \beta - 2 \Lambda M/\hbar \Omega.$ 

 The transfer  Hamiltonian  (\ref{6})  also undergoes a transformation.  Within the  accepted model, the second term in the expression for  the operator $"s" $ commutes with $ H_T .$ Thus $ H_T $ transformation reduces to substitution of renormalized coupling parameters $ \tilde \tau_{r\beta\sigma} $  for  $ \tau_{r \beta \sigma: }$   
\be
\tilde \tau_{r \beta \sigma} = \tau_{r \beta \sigma} X \equiv \tau_{r \beta \sigma} \exp \left [-\frac{\Lambda}{\hbar \Omega} (a^\dag - a) \right]  \label{7}
\ee 
 In addition to the Hamiltonian (\ref{1})-(\ref{5}) we assume that the vibrational mode is coupled to a thermal phonon bath and that this coupling is strong enough so that the phonon maintains its thermal equilibrium state throughout the process. 
 Consequently, the expectation value of the phonon operator $ X $ may be presented as follows  \cite{83}:
\be
\big< X \big> = \exp \left[- \left(\frac{\Lambda}{\hbar \Omega} \right)^2 \left(N_{ph} + \frac{1}{2}\right) \right]  \label{8}
\ee
where $ N_{ph} $ denotes the equilibrium phonon population.
 Here, we consider the low temperature regime where $ k \theta \ll \Lambda, \, \hbar \Omega, $ so $ \big< X \big> $ may be approximated by $ \ds \exp\left[-\frac{1}{2} \left( \frac{\Lambda}{\hbar \Omega} \right)^2 \right], $  which does not depend on temperature.  By replacing $ X $ by $ \big< X \big> $ we decouple electron and phonon subsystems. Then the  electron Green's function on the Keldysh contour  may be approximated as a product of the pure electronic term computed basing on the Hamiltonian 
 $ \tilde H = \tilde H_M + \tilde H_T \ (\tilde H_T $ being the transformed transfer Hamiltonian) and the Franck-Condon factor  \cite{37,38,84,85}:
\begin{align}
G_{i \sigma, j \sigma'} (t,t') &  
 \approx  - \frac{i}{\hbar}
\big < T_c d_{i\sigma} (t) d_{j\sigma'}^\dag (t') \big >_{\tilde H}
 \big<X (t) X^\dag (t') \big>
\nn\\ & \equiv
\tilde G_{i\sigma,j\sigma'} (t,t') K (t,t').  \label{9}
\end{align}
  This approximation is the most sensitive step in the accepted computational approach. It is inherent within the Born-Oppenheimer approximation and justified when a significant difference occurs between timescales characterizing electronic and vibrational dynamics. In the processes involving electron transport through molecules this difference in timescales is by no means obvious. However, it was shown that the Born-Oppenheimer approximation in the diabatic representation may be employed in analysis of such processes as well [86], and this is how the approximations (9) may be understood. 

The Fourier transform of the retarded Green's function for the electrons on the bridge may then be found in the form:
  \begin{widetext}
\be 
\tilde G_r^{-1} (E) =  \left [\ba{cccccc}
E - \tilde E_0 - \frac{i \Gamma_L}{2} & -\tilde\beta & 0 & 0 & \dots & 0  
 \\
-\tilde \beta & E- \tilde E_0 & -\tilde\beta & 0 & \dots & 0
\\
0  & -\tilde\beta & E- \tilde E_0 & - \tilde\beta & \dots & 0
\\
\dots & \dots & \dots & \dots & \dots & \dots
\\
0 & 0 & \dots & -\tilde\beta & E- \tilde E_0 & -\tilde\beta
\\
0 & 0 & \dots &  0 &- \tilde\beta & E - \tilde E_0 - \frac{i \Gamma_R}{2}
\\ 
\ea \right ] .  \label{10}
\ee
\end{widetext}
In this expression $(\gamma  = L,R):$
 \begin{align}
\Gamma_\gamma & = 2\pi \sum_{r,\sigma} |\tilde \tau_{r \gamma \sigma}|^2 \delta (E - \epsilon_{r \gamma \sigma} ) 
\nn\\  & = 
\exp \left [- \left(\frac{\Lambda}{\hbar \Omega} \right)^2 \right] 
2\pi \sum_{r,\sigma} \big| \tau_{r \gamma \sigma} \big|^2 \delta ( E - \epsilon_{r\gamma \sigma}) 
\nn \\ & \equiv
\Gamma_{\gamma 0} 
\exp \left [- \left(\frac{\Lambda}{\hbar \Omega} \right)^2 \right] 
 \label{11} .
\end{align} 
  Thus the coupling between the bridge and the electrodes  is reduced due to the effect of molecular vibrations, being  a manifestation of  the Franck-Condon physics typical to this model, which is further expressed by the dynamical term $ K(t,t') $ discussed below.
Within the wide band approximation, one may disregard dependences of $ \Gamma_{L,R} $ on $ E $ and treat these parameters as constants. In the following calculations we focus on a symmetrically coupled single-molecule junction: $ \Gamma_L = \Gamma_R = \Gamma. $  We remark that at the low temperatures considered here this symmetry is not violated by electron-phonon interactions.  

When gateway states are taken into consideration the equation for the retarded electron Green's function accepts the form:
\begin{widetext}
\be 
\tilde G_r^{-1} (E) =  \left [\ba{cccccc}
E - \tilde \epsilon - \frac{i \Gamma}{2} & -\tilde\delta & 0 & 0 & \dots & 0
 \\
-\tilde\delta & E- \tilde E_0 & -\tilde\beta & 0 & \dots & 0
\\
0  & -\tilde \beta & E- \tilde E_0 & - \tilde\beta & \dots & 0
\\
\dots & \dots & \dots & \dots & \dots & \dots
\\
0 & 0 & \dots & -\tilde\beta & E- \tilde E_0 & -\tilde\delta
\\
0 & 0 & \dots &  0 &- \tilde\delta & E - \tilde \epsilon - \frac{i \Gamma}{2}
\\ 
\ea \right ] .  \label{12}
\ee
\end{widetext}
This equation is derived assuming that the bridge chain is symmetrically coupled to the electrodes. Note that the on-site energy $ \tilde \epsilon $ and the coupling parameter $\tilde \delta $ are shifted due to electron-phonon interactions in the same way as $ E_0  $  and $ \beta.$  

Next consider the function K(t,t'). Since we assumed that the phonon subsystem maintains its thermal equilibrium, it can be evaluated to yield \cite{85}:
\be
K (t,t') = \exp[ - \Phi (t - t') ]  \label{13}
\ee
where
\begin{align} 
 \Phi (t' - t) = & \frac{\Lambda^2}{(\hbar \Omega)^2} \Big[ N_{ph} \big(1 - e^{i \Omega (t' - t)} \big) 
\nn\\ &
+ (N_{ph} + 1) \big (1 -  e^{-i \Omega (t' - t)} \big) \Big].  \label{14}
\end{align} 
Expanding $ K (t, t') $ in the terms of Bessel functions one obtains the following expression for the spectral  function matrix elements:
\begin{align}
A_{i\sigma, j\sigma'} =  i &  \sum_{m = \infty}^\infty L_m  \Big[ \tilde G_{i\sigma, j\sigma'}^<  (E - m \hbar \Omega) 
\nn\\    & \hspace{11mm}  -  
\tilde G_{i\sigma, j\sigma'}^>  (E + m \hbar \Omega) \Big] .  \label{15}
\end{align}
Here, the factors $ L_m $ have the form:
\begin{align}
L_m = & \exp \left[ \left(-\frac{\Lambda}{\hbar \Omega} \right)^2 \big(2N_{ph} + 1 \big) \right]   \exp \left[ \frac{m \hbar \Omega}{2 kT} \right]
\nn\\ & \times
I_m \left[ 2 \left(\frac{\lambda}{\hbar \Omega} \right)^2 \sqrt{N_{ph} (N_{ph} + 1) } \right]    \label{16}
\end{align}
and $ I_m (Z)  $ is the modified Bessel function of the $m $-th order. 

Specific manifestations of electron-vibron interactions in transport characteristics of molecular junctions are determined by three relevant energies. These are the coupling strengths of the electrodes to the molecular linker  expressed by
 $ \Gamma_0 ,$ the electron-phonon coupling parameter $ \Lambda $ and the thermal energies  $ k \theta_{L,R} \ ( \theta_{L,R} $ being the temperatures associated with the left and right electrodes). Here, we consider the situation when the system is weakly coupled $(\Gamma_0 \ll \Lambda,\hbar \Omega) $ and the characteristic temperatures are low $(k \theta_{L,R} \ll \Gamma_0). $         Also, we assume that electron-phonon coupling is rather weak $ (\Lambda < \hbar \Omega). $ 
 Under these conditions, one may expect a pronounced vibrational structure to appear in the electron transmission function which may significantly affect characteristics of thermoelectric transport.    Expanding  the Bessel functions in power  series and keeping first terms in these expansions we get: 
\be
 L_m \approx \exp \left[-\frac{\Lambda}{\hbar \Omega} \right]^2   \left( \frac{\Lambda}{\hbar \Omega} \right)^{2|m|} \frac{1}{|m |!},  \label{17}
\ee
 so, the expression (\ref{15}) may be reduced to: 
\be
A_{i\sigma, j\sigma'} = 2 \sum_{m = -\infty}^\infty L_m \tilde A_{i\sigma, j\sigma'} (E - m \hbar \Omega)).  \label{18}
\ee 

 The charge  current flowing through the symmetrically coupled junction then takes the form  \cite{87}: 
\begin{align}
I = &\frac{e}{4 \pi \hbar} \exp \left[ \frac{\Lambda^2}{(\hbar \Omega)^2} \right]
\int  dE 
\nn \\ & \times   Tr
\big\{ \big [ f_L(E) \Gamma_L - f_R(E) \Gamma_R \big] A(E) \big \}.  \label{19}
\end{align} 
Here, $ f_{L,R} (E) $ are Fermi distribution functions for electrodes. In the considered system where only first and last sites on the molecular  bridge are coupled to electrodes, each  $ N \times N $ transfer matrix $ \Gamma_{L,R} $ has a single nonzero element $ \Gamma_L^{1\sigma,1\sigma} = \Gamma_R^{N\sigma,N\sigma} = \Gamma. $ Using this feature and expression (\ref{15}) for spectral function, we may present the current $ I $ in Landauer  form:
\be
I = \frac{e}{\pi \hbar} \int dE \tau(E) \big[ f_L(E) - f_R (E) \big] \label{20}
\ee
where electron transmission function equals:
\be
\tau (E) = \frac{\Gamma^2}{4} \sum_{m = - \infty}^\infty P(m) \big | \tilde G_{1N} (E - m\hbar \Omega) \big|^2  \label{21}
\ee
and:
\be
P(m) = \frac{1}{|m|!} \left(\frac{\Lambda}{\hbar \Omega} \right)^{2|m|} 
 .  \label{22}
\ee

The obtained expression for the electron transmission describes the situation when an electron on the bridge with the initial energy $ E $ may absorb and/or emit phonons thus moving to a state with the energy $ E \pm m \hbar\Omega. $ In weakly coupled junctions $ (\Gamma_0, \Lambda \ll \hbar \Omega) $ 
 the  broadening of these states is sufficiently small for them to serve as transport channels for traveling electrons and the electron transmission may be roughly estimated as a sum of contributions from all these channels.

When the electrodes are kept at different temperatures, the on-site energies acquire corrections proportional to the thermal energy $ k \Delta \theta\ (k $ being the Boltzmann constant) making the Green's function matrix elements temperature-dependent  \cite{75,77}. These corrections may bring noticeable changes in the electron transmission provided that on-site energies take on values comparable with $ k \Delta \theta. $ However, when $ E_0 \gg k \Delta \theta $ which is typical for molecular junctions at sufficiently low temperatures, the effect of temperature on the Green's function becomes negligible. In further analysis, we omit these corrections thus making  the Green's function and electron transmission temperature independent.

As follows from Eqs. (\ref{10}), (\ref{12}) the retarded electron Green's function is represented by $ N \times N $ matrix. Solving Eq. (\ref{10}), one gets an explicit expression for the matrix element $ G_{1N} $ in the case of a simple chain \cite{79,88,89}:
\be
\tilde G_{1N} (E) = \frac{\tilde\beta^{N-1}}{\Delta_N (E,\Gamma)} .  \label{23}
\ee
Here, the determinant $\Delta_N (E, \Gamma ) $  is given by:
\begin{align}
\Delta_N (E, \Gamma ) =  \frac{1}{2^{N+1} \zeta}  &  \label{24}
\Big[(\lambda + \zeta)^{N-1} (\lambda + \zeta + i \Gamma)^2 
\nn\\ &
- (\lambda - \zeta)^{N-1} (\lambda - \zeta + i \Gamma)^2 \Big ]
\end{align}
where, $ \lambda = E - \tilde E_0,\ \zeta = \sqrt{\lambda^2 - 4 \tilde\beta^2}. $

When we take  into consideration the gateway states, the expression for $\tilde G_{1N} $ accepts the form:
\be
\tilde G_{1N} (E) = \frac{\tilde\delta^2 \tilde\beta^{N-3}}{\tilde \Delta_N (E,\Gamma)}   \label{25}
\ee
and the determinant $ \tilde \Delta_N (E,\Gamma) $ for $ N \geq 3 $ equals \cite{80}:
\begin{align}
 & \tilde \Delta_N (E,\Gamma) 
\nn \\ = &  
\Delta_N (E,\Gamma)  + (\alpha - \lambda)(\alpha + \lambda + i \Gamma) \Delta_{N -2} (E,0)
\nn\\ & +
\big[(\tilde\beta^2 -\tilde\delta^2)(\alpha + \lambda + i \Gamma) - (\alpha - \lambda)(\tilde\beta^2 +\tilde\delta^2) \big]
\nn\\  & \times
\Delta_{N-3} (E,0) - (\tilde\beta^4 - \tilde\delta^4) \Delta_{N-4}(E,0).   \label{26}
\end{align}
In this expression, $ \alpha = \tilde E_0 -  \tilde \epsilon,\ \Delta_N (E,\Gamma) $ is given  by Eq. (\ref{24}) and other determinants values are obtained from Eq. (\ref{24}) by putting $ \Gamma = 0. $ In particular, $ \Delta_0 (E,0) = 1 $ and $ \Delta_{-1} (E,0 ) = 0. $ 

In the following analysis we assume that the temperature of the right electrode is kept constant whereas the temperature of the left electrode varies. Also, we assume that $ \theta_L > \theta_R, $ so $  \Delta \theta = \theta_L - \theta_R > 0.  $  According to its definition, the thermocurrent is given by:
\be
I_{th} (V) = I (V, \theta_R, \Delta \theta) - I (V, \theta_R, \Delta \theta = 0)   \label{27} 
\ee
where $ V $ is the bias voltage applied across the junction and the Fermi functions $ f_{L,R} $  are computed for different leads temperatures and for different chemical potentials $ \mu_{L,R} $ of the leads. The chemical potentials are shifted with respect to each other by the bias voltage.   We suppose that $ V $ is symmetrically distributed between the electrodes, so $ \mu_{L,R} = \mu \mp eV/2,\ \mu $ being the chemical potential of electrodes in an unbiased system.  As the electron charge is negative, a positive bias  voltage shifts $ \mu_L $ above $\mu_R. $ Within the accepted approximations, electron transmission $ \tau (E) $ included in the integrand in the Eq.  (\ref{20}) is temperature independent and given by Eq. (\ref{21}). We apply Eqs. (\ref{18})-(\ref{26}) to analyze the dependence of  the  thermocurrent on the molecular bridge length.

\section{iii. Results and discussion}

 Consider first  the case where the molecular vibrations do not affect electron transport through the system. Then the expression for the electron transmission  takes  the simple form: 
\be
\tau (E) = \frac{\Gamma_0^2}{4} \big|  \tilde G_{1N} (E) \big |^2  .  \label{28}
\ee 
In the following computations concerning the bridge with gateway states we use   $ E_0 = - 4.47 eV,\ \epsilon = - 1.85 eV,\ \Gamma_0 = 2.85 eV,\ \delta = 1.27 eV $ and $ \beta = 2.28 eV. $ These are the  values which were derived for single-molecule junctions with gold leads and oligophenyl  bridges \cite{19}. For a simple tight-binding chain we put:  $ E_0 = - 4.6 eV,\ \beta = 2.2 eV $ and $  \Gamma_0 = 3 eV. $ We use two close sets of relevant parameters to better elucidate the effect of gateway states on the thermocurrent. 
In both cases, the HOMO appears to be located slightly below the chemical potential  of electrodes in the unbiased system  $ (\mu = 0) ,$ 
 and therefore,  serves as the primary channel for the
thermally induced transport.  Consequently, the charge carriers pushed through the system from the left (hot) to the right (cool) electrode by the thermal gradient $ \Delta \theta $ are holes, and the thermocurrent $ I_{th} $ through an unbiased junction takes on positive values.  Obviously, the thermocurrent increases with $ \Delta \theta. $ This dependence is linear for small $ \Delta \theta $ but it becomes superlinear at higher thermal bias, as presented in Fig. 2. 

\begin{figure}[t] 
\begin{center}
\includegraphics[width=7cm,height=5.65cm]{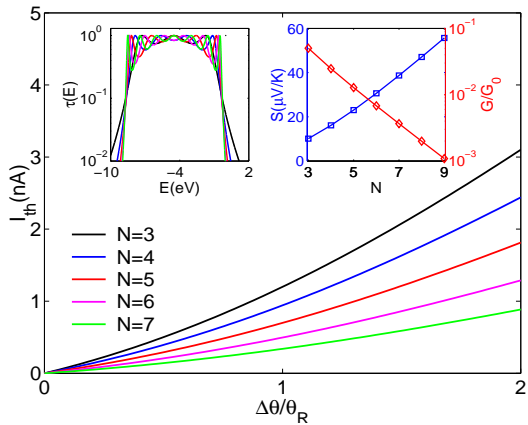} 
\caption{ Thermally excited current $ I_{th} $ in an unbiased molecular junction as a function of temperature for different lengths of the molecular bridge (different number of sites on the bridge chain).  Insets show the electron transmission for a simple chain plotted as a function of tunnel energy (left) and  the length dependences of the molecular conductance $ G $ and the thermopower  $S $(right). All curves are plotted for $ k \theta_R =  6 meV, \  \mu =0,\  E_0 = -4.6 eV,\  \beta = 2.2 eV,\  \Gamma_0 = 3eV. $
}
 \label{rateI}
\end{center}\end{figure}

Turning now to the bridge-length dependence of the thermocurrent, Figures 2 and 3 show that $ I_{th} $ decreases, for a given $ \Delta \theta, $ when the bridge length $ N $ increases. This may be contrasted with the observation \cite{13,15,16,17,18,19} that the Seebeck coefficient increases with  bridge length. 
 This apparent contradiction may be resolved by turning to the length dependence of the bridge conductance $ G. $ The latter falls very rapidly with increasing bridge length, and this fall is typically much more pronounced than the corresponding  increase in the thermopower $ S .$ An example of comparative behavior of these characteristics is displayed in the Fig. 2 (see right inset). The curves plotted in the inset are computed using the expressions  \cite{90}:
\be
G = \frac{e^2}{\pi \hbar} \tau (\mu) \equiv G_0 \tau (\mu)  , \label{29}
\ee
\be
S = \frac{\pi^2 k^2 \theta_R}{3e \tau (\mu)} \frac{\partial \tau (E)}{\partial E} \Big|_{E = \mu}.  \label{30}
\ee
These expressions satisfactory describe $ G $ and $ S $ at low temperatures provided that the transmission  $ \tau (E) $ smoothly varies in the close vicinity of the chemical potential  $ \mu:\  (|E - \mu| < k \theta_R). $ So, the weakening of $ I_{th} $ originates from the exponential fall of the molecular conductance, and it may appear simultaneously with the rise  of  the thermopower.

\begin{figure}[t] 
\begin{center}
\includegraphics[width=7cm,height=5.65cm]{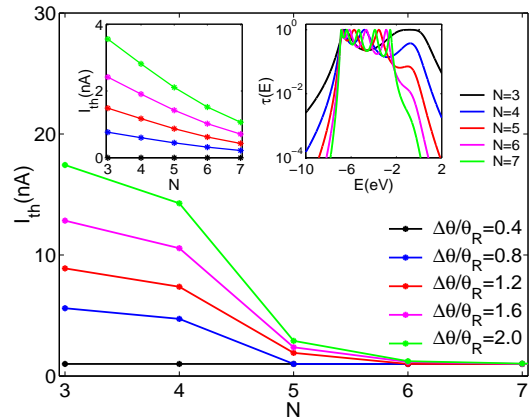} 
\caption{ The effect of gateaway states on the length-dependent thermally excited current for different values of $ \Delta \theta. $ Curves are plotted assuming   $ k \theta_R =  6 meV, \  \mu =0,\  E_0 = -4.47 eV,\  \beta = 1.27 eV,\  \Gamma_0 = 2.85eV,\ \delta = 2.28 eV. $ Insets show  $ I_{th} $ as a function of the bridge length for a simple chain computed using the same values for relevant parameters as in Fig. 2  (left) and the effect of gateaway states on the electron transmission function (right).
}
 \label{rateI}
\end{center}\end{figure}

Electron transport driven by a thermal gradient through an unbiased single-molecule junction  is dominated by a single HOMO/LUMO orbital. The dependence of thermoelectric transport characteristics (including $ I_{th}) $ on the molecular bridge  length results from the fact that the peak in $ \tau (E) $ associated with the transmitting becomes sharper and narrower as the length of the  molecular bridge increases. 
 The accepted values of relevant energies result in asymmetry of the model for no energy levels appear above the Fermi level. This asymmetry was deliberately introduced to make changes in the HOMO/LUMO  profile originating from variations in the bridge length more pronounced, as it happens with terminal transmission peaks within tight-binding models. However, we remark that the sharpening of the electron transmission peaks caused by the molecule lengthening is their inherent property. One may expect this effect to be manifested irrespectively of the specific model describing molecular bridge.

\begin{figure}[t] 
\begin{center}
\includegraphics[width=7cm,height=5.65cm]{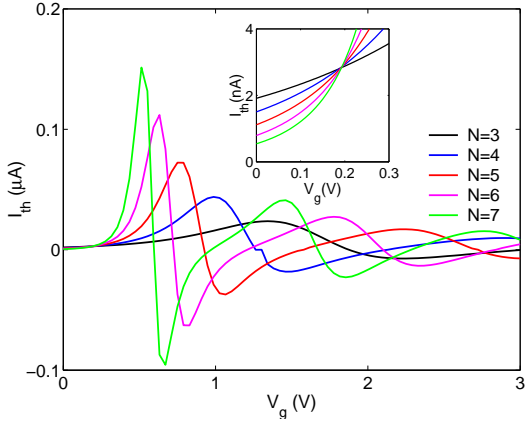} 
\caption{Thermocurrent  as a function of gate voltage $ V_g $  plotted at $ k \theta_R = 6 meV,\  \Delta \theta /\theta_R  = 1.4 $ for several different lengths of the molecular bridge. The inset shows {\color{Brown}  $I_{th} $ vs $ V_g $ plotted at small values of $ V_g. $} All remaining relevant parameters used in plotting the curves have the same values as those used in Fig. 2. 
}
 \label{rateI}
\end{center}\end{figure}

When gateway states are present,  significant changes in the profile  of  the HOMO/LUMO transmission peak may happen causing variations in  the length dependences of thermoelectric transport characteristics  \cite{19,80}. In the chosen model we may analyze changes in the thermocurrent originating from the effect of gateway states.                   As shown in Fig 3 (right inset), in this case the HOMO peak in the plot of $ \tau (E) $ versus $ E $ appears to be significantly broader than other resonance features. The peak broadening is  especially pronounced for relatively short bridges $(N \leq 5).$ The distorted HOMO profile is supposed to be responsible for nonlinear temperature dependences  of thermopower observed in experiments on molecular junctions [19]. In the present case, the same HOMO distortion causes a rapid fall of $ I_{th} $ occurring when the number of sites increases from $ N = 3 $ to $ N = 5 .$ For longer bridges, the distortion  of HOMO profile becomes less distinct, and the fall of $ I_{th} $ slows down. 
  On the contrary, in the case of a simple chain $ I_{th} $ decreases nearly uniformly as the chain lengthens.

 To further elucidate the nature of electron transport controlled by a thermal bias, we assume that the gate voltage $ V_g $ is applied to the junction, which shifts molecular bridge energy levels upwards along the energy scale.                      As follows from $ \tau(E) $ profiles shown in Figs. 2,3, at small $ V_g $ magnitudes the electron conduction remains low, and $ I_{th}$ takes on low values (see inset in Fig. 4). As $ V_g $ increases, transport channels associated with higher conductance come into play bringing a significant increase in the thermocurrent  magnitude. 
    When a certain molecular level approaches the electrodes chemical potential $ \mu $ from below, this level starts to serve as a transport channel for holes pushed by the thermal bias from the left (hot) electrode to the right (cool) one. Accordingly, the thermocurrent takes on positive values. When this level crosses $ E = \mu $ the holes flow becomes counterbalanced by the electrons flow  in the same direction, and $ I_{th} $ equals zero. When the molecular level is shifted slightly above $ E = \mu, $ it serves as a transport channel for electrons, so $ I_{th} $ accepts negative values. As a result, a derivative-like feature appears  in the $ I_{th} $ versus $ V_g $ plot. The electron transport induced by a thermal gradient applied across an electrically unbiased junction occurs solely on the condition that the molecular orbital serving as a transport channel is located in a close proximity of the electrodes chemical potential. 
Therefore,  when the molecular energy level is shifted farther upwards, away from $ E = \mu, $ it ceases to serve as a transport channel. 

\begin{figure}[t] 
\begin{center}
\includegraphics[width=7cm,height=5.65cm]{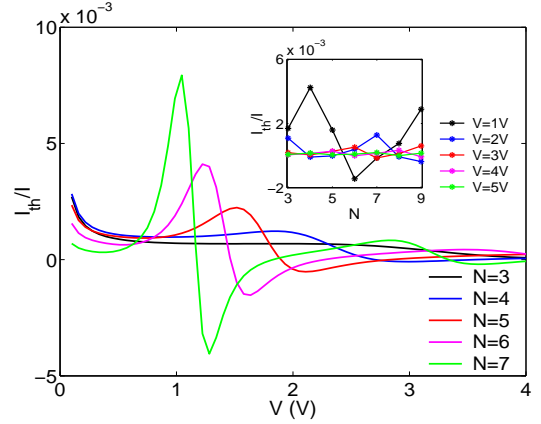} 
\caption{The ratio of thermocurrent $ I_{th} $ computed assuming that $ k\theta_R = 6 meV $ and $ \Delta \theta/\theta_R = 1.4 $ and charge current $ I $ flowing through the junction kept at a uniform temperature $(k\theta_L = k \theta_R = 6 meV) $as a function of bias voltage. Curves are plotted for several different lengths of the molecular bridge.  The inset shows $ I_{th}/I $ length dependences at several values of the bias voltage.
All remaining relevant parameters used in plotting the curves have the same values as those used in Fig. 2. 
}
 \label{rateI}
\end{center}\end{figure}	

As the gate voltage increases, another molecular energy level  may approach the chemical potential bringing another derivative-like feature into  the $ I_{th} $ versus $ V_g $ plot. The resulting  $ I_{th} (V_g) $ behavior for our model bridge is    shown in Fig. 4. 
 When the transmission peaks are well separated, the total number of derivative features is equal to the number of bridge states, namely the number of bridge sites.

When  $V_g = 0$ and a bias voltage $V $  is applied across the system, electron transport is simultaneously driven by electric and thermal biases. 
  The magnitude of $ I_{th} $	is determined by $ \Delta \theta $ regardless of the electrical bias  strength and polarity provided that the bias is sufficient for molecular orbitals associated with relatively high electron conductance to appear in the conduction window. 
  Its sign, however, depends on the voltage bias. Each time a molecular transmission channel crosses $ \mu, $ the sign of $ I_{th} $ changes, imparting an oscillatory contribution to the overall current. 
 The thermocurrent defined by Eq. (\ref{27}) represents a relatively small part of the total charge current $ I $ induced by the combined action of electrical and thermal driving forces even in weakly electrically biased molecular junctions.     	This is illustrated in the Fig. 5. 

\begin{figure}[t] 
\begin{center}
\includegraphics[width=7cm,height=5.65cm]{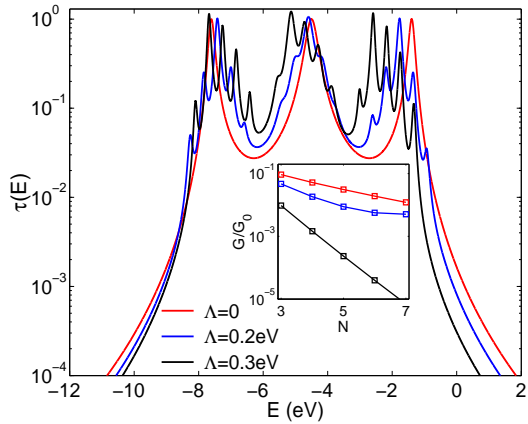} 
\caption{Electron transmission through a junction with a chain-like bridge affected by electron-phonon interactions. Inset shows the effect of these interactions on the molecular bridge conductance. Curves are plotted for $ N = 3,\  \Gamma_0 = 0.2 eV,\  \hbar \Omega = 0.42 eV ,\ M = 0.2 \Lambda . $ Remaining parameters take  on the same values as those used in Fig. 2.
}
 \label{rateI}  \end{center}\end{figure}
	
Similar oscillations of electrical conductance and thermopower were studied for multilevel quantum dots weakly coupled to electrodes \cite{59,91,92}.  Thermopower oscillations accompanying varying chemical potential of the electrodes were observed in single-molecule junctions  \cite{93}. It was shown that Coulomb interactions between electrons on a multilevel dot significantly affect oscillations of electron transport characteristics in these systems whereas in the present work we did not take these interactions into account. Nevertheless, we believe that the main reason for these oscillations and quasi-oscillations  of $ I_{th} $ described here is the same. These features appear when molecular orbitals (or quantum dot levels) cross  the boundaries of the conduction window. The latter may be created in different ways  and obviously different reasons may cause the shift of energy levels of the quantum dot and/or molecule.
	
				
	\begin{figure}[t] 
\begin{center}
\includegraphics[width=7cm,height=5.65cm]{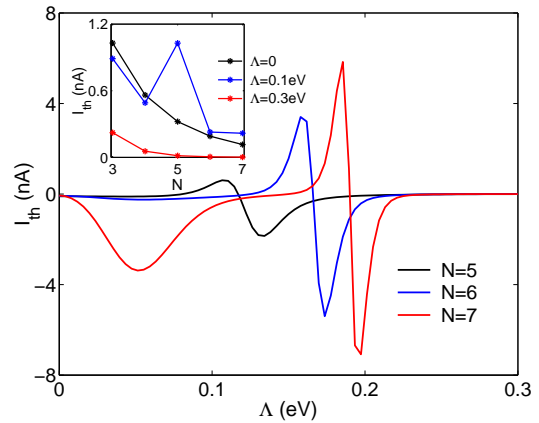} 
\caption{Thermocurrent in the unbiased molecular junction as a function of electron-vibron coupling strength plotted for different bridge lengths. Inset shows length-dependent $ I_{th} $ at different values of $ \Lambda. $ All curves are plotted assuming that $ k \theta_R = 6 meV,\  \Delta \theta/\theta_R =2,\  \Gamma_0 = 0.2 eV, \hbar \Omega = 0.42 eV,\  E_0 = - 4.6 eV,\  \beta = 2.2 eV, \  \mu = 0  ,\  M = 0.2 \Lambda. $
}
 \label{rateI}
\end{center}\end{figure}

As discussed before, in a weakly coupled system 
 the effect of vibronic interactions may be analyzed by considering the contribution of different vibronic levels to the transmission. This leads to appearance of vibronic structure in the transmission function (observed in inelastic tunneling spectroscopy) and to corresponding effect on the thermal conductance.  
  In Fig. 6 we display  the transmission function $ \tau (E) $  calculated using Eq. (\ref{21}) for a 3-site bridge $ (N = 3). $ 
	 One observes that in the presence of electron-vibron interactions each of the three electronic transmission peaks is replaced by a set of narrower peaks associated with the  vibronic levels. The center of each set is shifted with respect to the original peak position. This happens due to the polaronic shift of energy levels on the bridge. These changes in the electron transmission profile take place over the whole range of relevant values of the tunnel energy $ E, $ including the interval around $ E = \mu.$  As before, we assume that $ \mu =0. $   As shown in Fig. 6, the profile of $ \tau (E) $ near $ E = \mu $ may be significantly distorted. The $ \tau (E) $ profile varies depending on the electron-phonon coupling strengths $ \Lambda $  and $ M $ as well as on 	the number of bridge sites and the phonon mode frequency. 

These  variations are expected to be most strongly pronounced in the
 characteristics of thermally induced electron transport through unbiased molecular junctions since in this case the conduction window around $ E = \mu $ is very narrow. One may expect  alterations
 in molecular conduction and thermocurrent  behavior to appear. For example, the length-dependent molecular conductance through a simple chain may be partly suppressed   as $ \Lambda $  and $ M $  increase      as displayed in Fig. 6.  
 The thermocurrent itself shows changes in behavior   reflecting the effect of   electron-vibron coupling (see Fig. 7). One sees that $ I_{th} $ may display a minimum and/or a derivative-like feature at certain values of $ \Lambda $ provided that the molecular bridge includes five or more sites, that is the bridge  is sufficiently long. These features are signatures of vibronic levels appearing in the electron transmission.  When  electron-phonon interaction   becomes sufficiently large
 these features disappear. This happens because  an increase in the coupling strengths $ \Lambda, M $ is accompanied by the increase of the polaronic shift (assuming that $ \hbar \Omega $ remains fixed),   and by decrease in the parameter $\beta $ which narrows down the energy range where $ \tau (E) $ values are not too small.   As a result, all transmission peaks slide to the left, leaving solely smoothly falling tails within the vicinity of $ E = \mu. $  Also, one may observe  changes in   the
$ I_{th} $ dependences on the bridge length  (see inset)
which appear due to the effect of the vibrational mode.   We note that Eqs. (\ref{10}), (\ref{21}), (\ref{22}) were derived assuming a simple tight-binding model for the molecular linker. Nevertheless, these expressions give a reasonable approximation for the electron transmission which  is suitable for semiqualitative analysis of the response of thermocurrent to molecular vibrations.

\section{iv. conclusion}

In this work we have studied some aspects of steady thermoelectric transport through a single-molecule junction with a chain-like linker of an arbitrary length. Specifically, we focus on the thermocurrent $ I_{th}, $ defined as the change in the charge current at a given bias voltage V due to an imposed temperature difference between the two leads. To compute $ I_{th}, $ we model the bridge as a tight binding chain of identical sites. We have highlighted several characteristics properties of the thermocurrent:

(a) In unbiased systems characterized by tunneling conductance $ I_{th} $decreases with increasing bridge length. This behavior is caused by the exponential fall of the bridge conductance with increasing bridge length. 

 (b)  The length dependence of $ I_{th} $ may be significantly affected by the profile of HOMO/LUMO peak in the electron transmission function. Specifically, the particular HOMO profile caused by gateaway states associated with terminal sites on the molecular bridge may significantly change behavior of the length-dependent thermocurrent. Similar effect of gateaway states on the thermopower of single molecule junctions was observed  and analyzed in earlier works [19,80]. We have shown that $ I_{th}$ may experience the gateaway states influence within a wide range of temperature variations. 

(c) When gate or bias potentials bring the system closer to resonance transmission  $ I_{th}, $ and its bridge length dependence are affected by the energy dependent variations in the electron transmission profile   near the electrodes Fermi energy.   
    In a gated/biased molecular junction the thermocurrent changes sign several times, as the  voltage increases. The profile $ I_{th}  (V_g ) $ or $ I_{th} (V) $ looks like a sequence of peaks and dips. The number of peaks in this picture corresponds to the number of conducting bridge states, which in the  considered model reflects the number of sites on the bridge chain. This change in sign indicates the change of charge carriers (electron/holes) involved in the transport process. It happens when a certain molecular orbital crosses the boundary of conduction window when either the bias or gate voltage is increasing. These peaks become more pronounced as the temperature gradient   rises. However, at sufficiently strongly gated/biased system when all molecular orbitals cross the boundary $ E = \mu $ or are situated inside the conduction window $ I_{th} $ becomes zero. 

(d) Within our model we have  analyzed the effect of molecular vibrations on the thermocurrent. In particular, we have found that in weakly coupled molecular junctions $ (\Lambda,\Gamma_0 \ll \hbar\Omega) $ at sufficiently low temperatures $(\Lambda,\Gamma_0 \gg k\theta)$ electron interactions with vibrational modes may qualitatively change the length dependences of $ I_{th} .$   Specifically, it was shown that $ I_{th} $ may display dips and/or derivative-like features at certain values of electron-phonon coupling parameters. These features may appear in system with sufficiently long molecular bridges $(N \geq 5),$ at sufficiently weak electron-phonon coupling  $(\Lambda < \hbar\Omega) $. They are signatures of vibronic levels occurring in the electron transmission. 

The system used in the present analysis is simple, and more detailed models, both regarding the molecular electronic structure  and its coupling to the environment are required for quantitative calculations of thermocurrent through long molecular bridges. Nevertheless, we believe that the results presented and discussed here capture some essential physics and may be helpful in studies of thermoelectric transport through tailored nanoscale systems. \vspace{2mm}

{\bf Acknowledgments:} We thank G. M. Zimbovsky for help with the manuscript preparation. This work was supported by the US NSF-DMR-PREM 1523463 and the U.S. NSF Grant No. CHE1665291.

 \end{document}